\documentstyle[12pt]{article}
\newcommand{\dss}{D^{\ast \ast}}
\input epsf
\begin{document}
\title{\bf Strong Decays of Hybrid Mesons \\ from the Heavy Quark
Expansion of QCD}

\author{Philip R. Page\thanks{prp@jlab.org. Work done in collaboration
with Eric S. Swanson and Adam P. Szczepaniak. Contribution to 
the Seventh International Conference on Hadron Spectroscopy
(HADRON '97), Brookhaven, August 1997. I acknowledge a Lindemann
Fellowship from the English Speaking Union.}$^*$ \\ \\
\it\small Theory Group, Thomas Jefferson National Accelerator Facility,\\
\it\small 12000 Jefferson Avenue, Newport News, VA 23606, USA }
\date{}
\maketitle

\begin{abstract}
We calculate the strong decays of hybrid mesons to
conventional mesons for all the lowest lying $J^{PC}$ hybrids of
flavour $u\bar{u},\; d\bar{d},\;
s\bar{s},\; c\bar{c}$ and $b\bar{b}$. A decay operator 
developed from the heavy quark expansion of
quantum chromodynamics is employed. We show that the selection rule that 
hybrid mesons do not decay to identical S--wave mesons, found in 
other models, is preserved. We predict decays of charmonium
hybrids, discuss decays of $J^{PC}=1^{-+}$ exotic isovector hybrids
of various masses, and interpret the $\pi(1800)$ as a hybrid meson.
\end{abstract}

\vspace{1cm}

As the Standard Model has successively been confirmed in recent years,
one remaining area of ignorance is the interactions of the strong
gluonic degrees of freedom. Particularly, quark--antiquark bound states where there is an explicit excitation of the
gluon field of QCD, called ``hybrid mesons'', have been predicted to
exist, but have so far not been uncovered unambiguously in
experiment. 

In this talk we outline a new model for the decay of hybrid mesons
into conventional mesons \cite{swanson97} to be reported in detail elsewhere \cite{page98er}.

Our description of hybrid mesons and conventional mesons
is that of the non--relativistic flux--tube model of Isgur and
Paton. We consider a connected decay of an initial hybrid A into final
state mesons B and C. Isgur, Kokoski and Paton proposed a decay
operator motivated from the strong coupling limit of the hamiltonian
formulation of lattice gauge theory, which we shall refer to as the
``IKP model'' \cite{kokoski85}. 
We propose a decay operator motivated from the heavy quark
expansion of QCD in Coulomb gauge \cite{swanson97}.

We uncover the selection rule that hybrid meson decays to two identical
angular momentum $L=0$  mesons vanish if the final state meson wave functions are
identical. This is also found in the IKP model and has recently been
proved to arise independent of detailed models \cite{page97sel2}.
However, in our model, the selection rule remains true even 
for mesons with general $L$ \cite{swanson97}.

We present some of the highlights of a calculation of the strong decays of hybrid mesons to
conventional mesons for all the lowest lying $J^{PC}$ hybrids of
flavour $u\bar{u},\; d\bar{d},\;
s\bar{s},\; c\bar{c}$ and $b\bar{b}$. 
If the hybrid decays differs from that of the IKP
model, we are able to ascertain model--dependence of
predictions and hence can apply some caution to results.  
In areas where the two models coincide they
provide robust predictions. 

Caution has to be applied to the overall normalization of decays. 
The IKP model predictions calculated here can be up to 2 times larger than
quoted, due to uncertainties in the pair creation strength
\cite{page98er} and correspondingly, our model's predictions, since its
normalization is fixed relative to the IKP model.

\begin{figure}[b!]
\label{figccbar}
\begin{center}
\leavevmode
\hbox{\epsfxsize=4 in}
\epsfbox{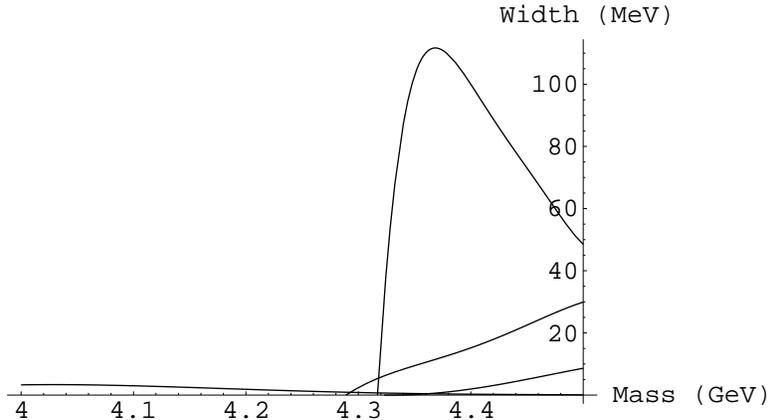}
\vspace{10pt}
\caption{Dominant partial widths of a $1^{-+}\; c\bar{c}$ hybrid at various
masses. The partial widths to $\dss(1^{+}_H)D,\; \dss(1^{+}_L)D,\; 
\dss(2^{+})D$ and $D^{\ast}D$ correspond to the highest to the lowest intersections
with the vertical axis. $\dss({1^{+}_L})$ and
$\dss({1^{+}_H})$ (denoted $D_1(2420)$ in the PDG) are the low and high mass $1^+$ states respectively. } 
\end{center}
\end{figure}

We firstly discuss charmonium hybrids. 
The widths of charmonium hybrids are suppressed below
the $\dss D$ threshold ($\approx 4.3$ GeV), where only $D^{\ast}D$ and $D^{\ast}_s D_s$
modes are allowed by the selection rule, since these are the only open charm combinations
where the two final state wave functions are sufficiently different.
However, when
states are allowed to become more massive than the $\dss D$ threshold,
the total widths increase drastically (see Fig. 1)
to $15 - 160$ MeV for 4.4 GeV hybrids. 

Hybrids often have exotic quantum numbers not found for conventional
mesons, e.g. $J^{PC} = 1^{-+}$ and $2^{+-}$. Exotic $J^{PC}$ immediately
identifies the state as not being a conventional meson.
The 4.4 GeV $c\bar{c}\; 2^{+-}$ exotic has a small width of $15$ MeV
in our model, but not in the IKP model. Fig. 1 indicates widths for
the $1^{-+}$ exotic.

\begin{figure}[b!]
\label{figrho}
\begin{center}
\leavevmode
\hbox{\epsfxsize=4 in}
\epsfbox{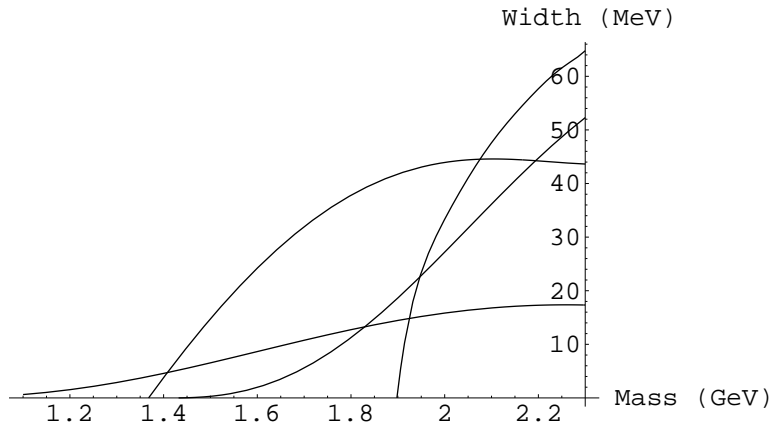}
\vspace{10pt}
\caption{Dominant partial widths of a $1^{-+}$ isovector hybrid at various
masses. The partial widths to $K_1(1400)K,\; \eta(1295)\pi,\; b_1\pi$
and $\rho\pi$ correspond to the highest to the lowest intersections
with the vertical axis.} 
\end{center}
\end{figure}

We now indicate results
for the $u,d$ quark isovector $J^{PC} = 1^{-+}$ hybrid. 
Our expectations for the widths of a  $1^{-+}$
of mass 2.0 GeV are (in MeV)
\begin{tabbing} \label{bnl2}
XXXXXXX\=XXX\=XXXXXXX\=XXXXXX\=XXX\=XXXXXX\=XXX\=XXX\=XXXXXXX\=XXX\kill 
\>$b_1\pi$\>$K_1(1400)K$\>$\eta(1295)\pi$\>$\rho\pi$\>$\rho(1450)\pi$\>$f_1\pi$\>$a_1\eta$\>$K_1(1270)K$\\
Our model\>40\>30\>30\>20\>10\>10\>10 \>10 \\
IKP model\>60\>80\>20\>20\>10\>40\>10\>20\\
\end{tabbing}
where we have neglected $K^{\ast} K,\; f_2\pi,\; \pi(1300)\eta,\;
K(1460)K$ and $K^{\ast}(1410)K$ which are predicted at $\leq 5$ MeV in both models,
and $\eta\pi,\; \eta^{'}\pi,\; \rho\omega,\; a_2\eta$ and
$K_2^{\ast}(1430)K$ with are 0 MeV
in both models. Because of the substantial phase space
available and the selection rule, $P+S$ channels are dominant.
Our model has several modes which are suppressed relative to the IKP
model. In addition to the important $b_1\pi$ channel, $K_1(1400)K$
emerges as a prominent channel.

For a 1.6 GeV state the widths are
\begin{tabbing}\label{bnl1}
XXXXXXXXXX\=XXXXX\=XXXXX\=XXXXX\=XXXXXXXX\=XXXXX\=XXX\kill 
\>$b_1\pi$\>$\rho\pi$\>$f_1\pi$\>$\eta(1295)\pi$\>$K^{\ast}K$\\
Our model\>20\>10\>5 \>2\>1\>MeV\\
IKP model\>60\>10\>10\>1\>0\>MeV\\
\end{tabbing}
where both models predict $\eta\pi,\; \eta^{'}\pi,\; \rho\omega$ and
$f_2\pi$ widths to be 0 MeV.
Superficially, the main effect of our model is to make the $P+S$ modes
of a more 
similar size to the $S+S$ modes than they are in the IKP model. 
We emphasize the importance of
searching for the hybrid in $\rho\pi$, as well as in the $b_1\pi$ and
$f_1\pi$ channels qualitatively preferred by the selection rule. Also, both models concur that $b_1\pi$ should primarily be
focused apon. 

The decay modes in
Fig. 2 demonstrate significant dependence on the hybrid mass.

Ref. \cite{swanson97} contains a calculation of the widths 
of the $\pi(1800)$ in our model which include below threshold
decays to $K^{\ast}_0(1430)K$ of 85 MeV.\protect\footnote{Some of the $K^{\ast}_0(1430)K$ mode predicted
in our model is expected to couple to $f_0(980)\pi$ via
$K^{\ast}_0(1430)K\rightarrow (K\pi)K\rightarrow f_0(980)\pi$
final state interactions, so that our model estimate is actually less than 85 MeV.} 
It is useful to correlate the decay modes
to experimentally known ratios. Specifically, using the 
VES experimental branching ratios\protect\footnote{The experimentally
measured $KK\pi$ ($\pi\pi$) in S--wave is assumed to arise solely
from $K^{\ast}_0(1430)K$ ($f_0(1370)$).} \cite{zaitsev95} and correcting
for decays of particles into the specific channels observed by VES
\cite{page97jpsi}, we obtain

\begin{tabbing} \label{bnl2}
XXXXXXXXX\=XXXXXXXX\=XXXXXXX\=XXXXX\=XXXXX\=XXXXXXX\=XXXXX\kill 
\>$K^{\ast}_0(1430)K$\>$f_0(1370)\pi$\>$\rho\pi$\>$K^{\ast}K$\>$\rho\omega$\>$f_1\pi$\\
Experiment\>$1.0\pm 0.3$ \>$0.9\pm 0.3$\> $<0.36$\> $<0.06$\>$0.4\pm 0.2$\>small\\
Our model\>$<0.7$\>0.6\>0.31\>0.05\>0\> 0.01\\
\end{tabbing}
where we have scaled our model widths evaluated in ref. \cite{swanson97} by a common factor to
compare to the experimental ratios deduced in ref. \cite{page97jpsi}. 
The correspondence is spectacular. Experimentally, significant
couplings to $f_0(980)\pi$ and $a_0(980)\eta$ are also observed
\cite{zaitsev95}, which should arise due to final state interactions
in our model.
We emphasize that although $\rho\pi$ is suppressed in the data, we
expect the resonance to have a non--negligible coupling to this channel.

One inconsistency with VES data appears to be the $\rho\omega$ mode. It is
significant that the resonance in $\rho\omega$ has a mass of $1.732\pm
0.01$ GeV \cite{amelin97}, shifted downward from the usual $\pi(1800)$
mass parameters. There are also indications of the presence of a 
broad $0^{-+}$ wave \cite{amelin97}. This may signal the presence of
the $u,d$ quark conventional meson expected in this mass region, removing the
apparent inconsistency with the hybrid interpretation of $\pi(1800)$.

\end{document}